\documentclass[11pt]{article}
 \pdfoutput=1
\usepackage[pdftex]{graphicx}
\usepackage{epsfig}
\usepackage{cite}
\newcommand{\AmS}{{\protect\the\textfont
  A\kern-.1667em\lower.5ex\hbox{M}\kern-.125emS}}
\newcommand{\ba}{\begin{array}}
\newcommand{\ea}{\end{array}}

\def\beq{\begin{equation}}
\def\eeq{\end{equation}}

\def\bea{\begin{eqnarray}}
\def\eea{\end{eqnarray}}

\def\hg{{\rm h}}
\def\ct{\cite}
\def\vg{{\rm v}}

\def\beq{\begin{equation}}   
\def\eeq{\end{equation}}

\def\bea{\begin{eqnarray}}
\def\eea{\end{eqnarray}}

\begin{document}
\begin{titlepage}

\begin{flushright}
CERN-TH-2018-103\\
 IFIC/18-29, FTUV-18-08-03\\
\end{flushright}

\begin{center}
\vspace{2.7cm}
{\Large{\bf 
 Signatures of new physics 
 versus the 
ridge phenomenon\\ 
\vskip 0.2cm
in hadron-hadron collisions at the LHC}}
\end{center}

\vspace{1cm}

\begin{center}

{\bf Miguel-Angel 
Sanchis-Lozano$^{\rm a,}$\footnote{Email 
address:Miguel.Angel.Sanchis@ific.uv.es}
and 
Edward K. Sarkisyan-Grinbaum$^{\rm b,c,}$\footnote{Email address: 
Edward.Sarkisyan-Grinbaum@cern.ch}
\vspace{1.5cm}\\
\it 

\it $^{\rm a}$ Instituto de F\'{\i}sica
Corpuscular (IFIC) and Departamento de F\'{\i}sica Te\'orica \\
\it Centro Mixto Universitat de Val\`encia-CSIC, 
Dr. Moliner 50, E-46100 Burjassot, Valencia, Spain
\\ 
$^{\rm b}$ Experimental Physics Department, CERN, 1211 Geneva 23, 
Switzerland\\
$^{\rm c}$ Department of Physics, The University of Texas at Arlington, 
TX 
76019, USA}

\end{center}

\vspace{0.5cm}

\begin{abstract}
 In this paper, we consider the possibility that a new stage of matter, 
stemming from hidden/dark sectors beyond the Standard Model, to be formed in 
$pp$ collisions at the LHC, can significantly modify the correlations 
among final-state particles.
 In particular, two-particle azimuthal correlations are studied by means of a Fourier 
series sensitive to the near-side ridge effect while assuming that hidden/dark 
particles decay on top of the conventional parton shower.
  Then, new (fractional) harmonic terms should be included in the Fourier 
analysis of the azimuthal anisotropies, encoding the hypothetical new 
physics contribution enabling its detection in a complementary way to
 other signatures.
 %
 
\end{abstract}
\begin{center}


\end{center}

\end{titlepage}


The interest of discovering new physics (NP) beyond the Standard Model (SM) at the LHC is out of doubt. Along
the last decades, many and distinct strategies have been put forward, most of them based on signatures in the 
transverse plane with respect to the beams axis like mono-jets, missing transverse energy, displaced vertices and so on. On the other hand, 
other kind of rather ``diffuse'' signals have been examined 
in the literature, e.g. \cite{Kang:2008ea,Harnik:2008ax}, 
featuring the whole event (multiplicity distribution and moments, event 
shape variables, underlying event etc.)
as a key signature of NP. For instance, the so-called ``soft bomb'' scenario \cite{Knapen:2016hky} is
characterized by high multiplicity events with nearly 
spherically distributed soft SM particles, and a large amount of missing transverse energy. 
In particular, a strongly coupled hidden/dark sector
could lead to large angle emission of partons carrying a non-negligible amount of momentum,  
yielding a rather isotropic distribution of 
final-state particles, all sharing a similar amount of energy. Notice, however, that a
likely complicated hidden sector (HS) beyond the SM may have limited
observable effects at colliders, making 
hard the detection from SM background and especially
the discimination among different models \cite{Cohen:2017pzm,Han:2007ae}. Thereby, alternative signatures,
as proposed in this work, should be considered 
 as complementary to other search strategies 
 as 
 discussed 
 in \cite{Knapen:2016hky}.    

Indeed, as is well known, (pseudo)rapidity and azimuthal particle 
correlations provide a crucial insight into the underlying mechanism of
particle production (see \cite{book} for a review). Moreover, from general arguments based
on causality, long-range correlations should have the origin at very early times after the collision. 
Therefore, if the parton shower were to be altered by the presence of a non-conventional state of matter,  
final-state particle correlations should be sensitive to it 
\cite{Sanchis-Lozano:2009}. 

The two-particle correlation function is 
 often
 defined in pseudorapidity and azimuthal 
 space as  
 \ct{ridge-pA}
 \begin{equation}\label{eq:corexp} 
C(\Delta \eta, \Delta \phi)= 
\frac{S(\Delta \eta,\Delta \phi)}{B(\Delta \eta, \Delta \phi)}\,\, .
 \end{equation} 
Here, $S$  
denotes the signal 
distribution built with particle pairs 
from the same event while 
$B$ stays for the background distribution constructed by particle pairs 
taken from  
 different events;
  $\Delta \eta=\eta_1-\eta_2$ and $\Delta 
\phi=\phi_1-\phi_2$ denote, 
respectively, the 
pseudorapidity and
 azimuthal differences of particles 
 $1$ and $2$, the indexes 
 labelling  
 the trigger and associate particles, respectively.

 Typically, a complex structure 
 of the correlation function 
 is observed, 
 in particular, 
an 
enhancement of the two-particle
correlations is 
found at $\Delta \phi \simeq 0$  in heavy-ion collisions
\cite{CollEff-rev}. Because of its 
extended longitudinal (pseudorapidity) shape,
as seen in the $\Delta \eta$--$\Delta \phi$ plot, it
is referred to as the (near-side) ridge. 
One-dimensional correlation functions $C(\Delta \phi)$
are obtained from Eq.~(\ref{eq:corexp}) by integration over pseudorapidity 
along
the range $2 < |\Delta \eta| < 5$ to focus
on long-range correlations.

The observed azimuthal anisotropy in heavy-ion collisions is commonly analysed by means of a Fourier decomposition: 
\begin{equation}\label{eq:cold}
C(\Delta \phi)\ \sim 1 + 2\sum_{n=1}^{\infty}V_n\cos{(n\Delta \phi)}\ ,
\end{equation}
where the coefficients $V_n$ are supposed to factorize as the product of the coefficients of the
equivalent Fourier expansion of two single-particle densities. When applied to heavy-ion collisions, the 
different terms in the series of Eq.~(\ref{eq:cold}) find a ``natural'' 
interpretation 
according to a hydrodynamical model
describing the very hot and dense matter resulting from the collision.    
In practice, up to five or six Fourier terms are taken into account in the analysis of the experimental data. 

Remarkably, similar long-range ridge structures show up in proton-nucleus 
\ct{ridge-pA,ridge-pp-pA-ATLAS}
 and even proton-proton \ct{ridge-pp-pA-ATLAS,ridge-pp} collisions, 
 under several conditions on events, like high multiplicity and a given 
transverse momentum range
of charged particles. The interpretation of a positive $V_2$ in these small systems is currently highly debated,
and different observables have been proposed to probe new dynamical effects related to
large hadronic densities \cite{Mangano:2017plv}.

\begin{figure}[ht!]
\vspace*{-1.5cm}
\begin{center}
\includegraphics[scale=0.34]{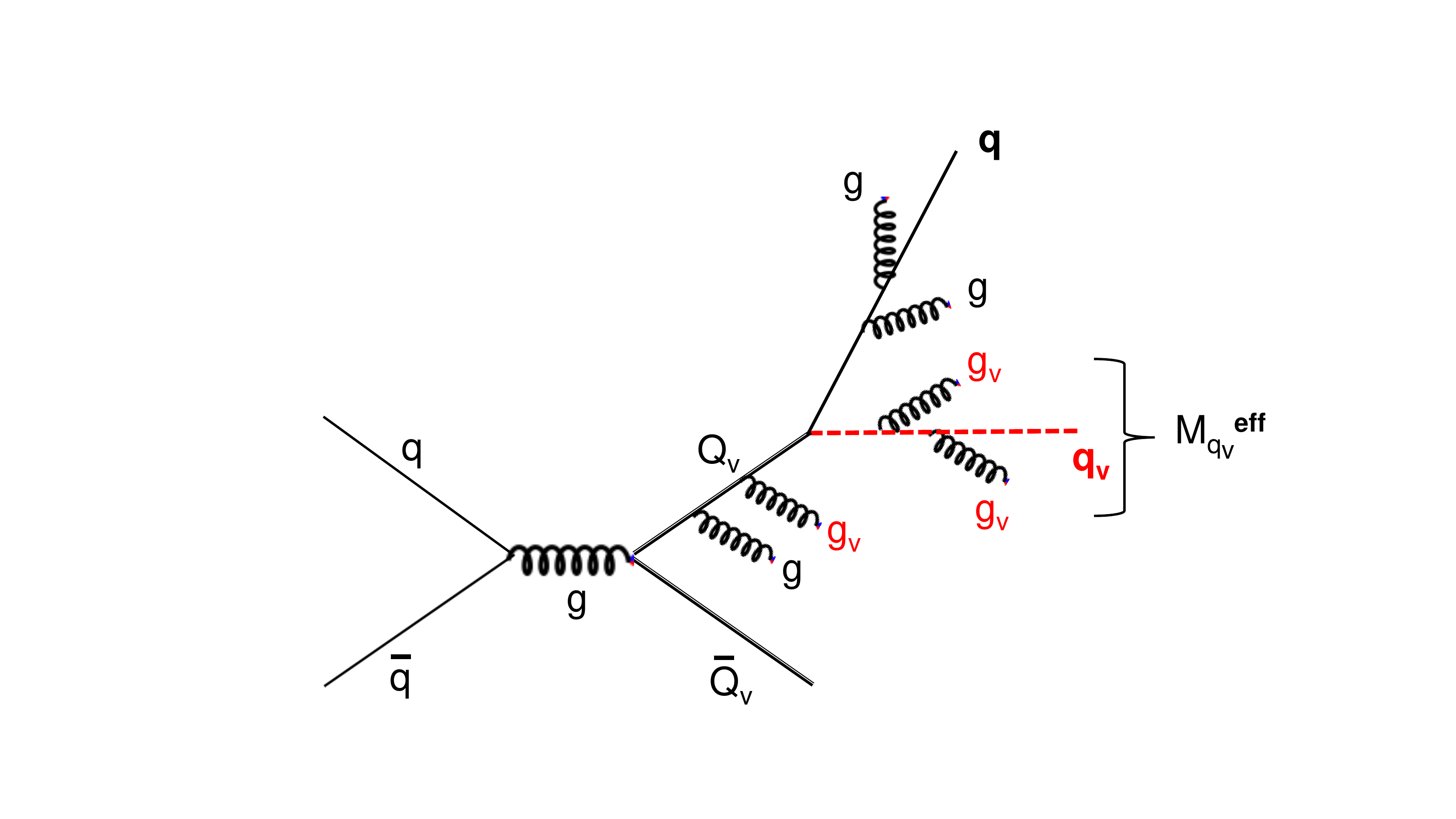}
\vspace*{-0.5cm}
\caption{Pair production 
via $q\bar{q}$ fusion  
of a pair of mediators ($Q_\vg\bar{Q}_\vg$) bearing both
SM and hidden charges. 
The decay $Q_\vg \to q+q_\vg$ can originate 
SM and hidden cascades from gluon and v-gluon emission.}
\label{fig:1}
\end{center}
\vspace*{-0.4cm}
\end{figure}

In this 
 paper 
 we furthermore consider that 
hidden particles and states, stemming from Hidden Valley (HV) models 
\cite{Strassler:2006im}, 
can be formed at primary interactions in very high energy $pp$ 
collisions. Generically, 
a HV model consists of three sectors:  (i) a HS containing v-particles 
charged under a valley group $G_\vg$ but 
blind to
the SM interactions, (ii) a visible sector including SM particles 
charged under the SM group $G_{\rm SM}$ but neutral under $G_\vg$, and
(iii) mediators connecting both visible and hidden sectors. Usually, the 
masses of the hidden sector particles are assumed to lie 
below the electroweak scale while the mediators may have TeV-scale masses.
The simplest possibility for $G_\vg$ is a QCD-like scenario, with a strong 
(running) coupling constant $\alpha_\vg$ and confinement scale 
$\Lambda_\vg$. The SM sector could feebly couple
to the HS (and the equivalent hadronic v-particles and states)
via a neutral $Z'$ or via heavy particles bearing both $G_{\rm SM}$ and 
$G_\vg$ 
charges. 
Here, we consider the latter possibility along the lines of the Monte Carlo (MC) study 
using PYTHIA \cite{Carloni:2010tw}, where the hidden shower is controlled basically by two parameters:  
the coupling strength $\alpha_\vg$ 
assumed to be a constant (i.e. no running is considered) and the lower cut-off scale set equal to 
0.4 GeV as by default in QCD showers, consistent with a low hidden confinement scale 
$\Lambda_\vg$.  
Such a simplified picture is compatible with the expected walking behaviour requiring a strong coupling 
over a large energy window along the showering
before reaching $\Lambda_\vg$, 
thereby yielding a large number of hidden partons and 
final-state particles. 
At the end, the energy from the primary interaction is democratically shared by soft final-state SM particles, while no classical jet structure is expected, thereby adapting quite well to a soft-bomb scenario. 
As commented, the ultimate goal in this paper 
is to show that long-range azimuthal correlations among final-state particles 
should emerge as a consequence of such 
kind of scenario.

Focusing only on the particle content relevant to 
 the study presented here, 
 we collectively denote by $Q_\vg$ the (spin 1/2) hidden partners of the 
SM 
quarks, 
charged under both $G_\vg$ and $G_{\rm SM}$, while $g_\vg$ and $q_\vg$ 
stand
for the v-gluon and (spin 0) v-quark only charged under the $G_\vg$ hidden 
group 
\footnote{
 The
 notation 
 used follow  that 
 of 
 \cite{Carloni:2010tw} 
 for
 the HS 
  in the PYTHIA 8 MC generator.}, respectively.

Special mention deserves the unparticle scenario, which can be viewed as a special
case of HV models. Let us recall that, from a phenomenological point of
 view, an unparticle 
 \cite{Georgi:2007ek}
 does not have a fixed 
 invariant mass squared
but instead a continuous mass spectrum.  As pointed out in the literature
(see e.g. \cite{Cheung:2008xu}), 
direct detection of unparticle stuff
at colliders should rely on peculiar missing energy distributions.
Unparticle production influence on particle correlations 
would become another useful tool to study such scenario, as 
   shown 
 in this work.

For some parameter values of HV models,  hidden particles
could promptly decay back into SM particles, 
altering the subsequent conventional parton shower \cite{Strassler:2008fv}
and yielding (among others \cite{Cohen:2017pzm}) observable consequences, 
e.g. extremely long-range
correlations especially in azimuthal space \cite{Sanchis-Lozano:2018wpz}. In this paper 
we do not enter into details about specific models but limit ourselves to general features associated with
the production of very massive objects on top of the parton shower
and their observable consequences, mainly from kinematic constraints. 

Our analysis focuses on $Q_\vg$ pair-production 
 via $gg$ or $q\bar{q}$ fusion (see Fig.~\ref{fig:1}), 
 subsequently
decaying into a v-quark and a SM quark: $Q_\vg \to q_\vg+q+X$, where $X$ 
stands for
an ensemble of radiated gluons and v-gluons which, in turn, will originate 
visible and hidden parton cascades.
Note that a very massive $Q_\vg$ would be produced at a rather
{\em low velocity} during the primary parton-parton interaction in $pp$ collisions at the LHC. In fact,  
assuming that the centre-of-mass subenergy of the parton-parton interaction
is of the order
of or higher than twice magnitude of the mediator mass ($\sqrt{\hat{s}} 
\ge 2M_{Q_\vg}$), then
$Q_\vg$ states can be on-shell pair-produced. Moreover, all 
(either SM or hidden) particles 
stemming from its decay should have access to a limited energy due to v-gluon 
radiation. In sum, 
final particles would ``democratically'' share the centre-of-mass 
energy released in the primary collision and rather 
soft and diffuse signatures are expected. 

Below we use velocity of very heavy hidden sources for kinematic estimates 
involving angular distributions \footnote{Velocity may become
a well-defined physical and meaningful quantity when dealing with heavy 
particles \cite{Neubert:1993mb}.}. Moreover,  
we assume an isotropic parton emission in the hidden particle rest frame, 
coming out from the primary interaction, slightly boosted 
in the laboratory reference frame due to 
 a non-relativistic velocity of the above-mentioned hidden source. We 
 consider that
 this assumption provides 
the essential framework for our estimates and conclusions.

In the hidden source $Q_\vg$ rest frame, the product of the velocity 
$v_\hg$ and 
the Lorentz 
factor $\gamma_\hg = \left(1-v_\hg^2\right)^{-1/2}$ of the fragmenting 
v-quark
is roughly given by
\begin{equation}\label{eq:vgamma}
v_\hg \gamma_\hg=\frac{M_{Q_\vg}^2-M_{q_\vg}^{{\rm 
eff}\,2}}{2M_{Q_\vg}M_{q_\vg}^{\rm 
eff}} 
\,\, ,
\end{equation}
where the bare $q$-mass was set equal to zero. The 
effective v-quark invariant mass, denoted as $M_{q_\vg}^{\rm eff}$, is 
defined in a similar way as 
in conventional QCD jets, i.e.
\beq\label{eq:invmass}
M_{q_\vg}^{\rm eff} = \sqrt{(\sum_{j} E_j)^2-(\sum_{j} \vec{p}_j)^2} \,\, 
,
\eeq
where $E_j$ and $\vec{p}_j$ stands for the energy and three-momentum 
of the v-gluons 
 emitted by the fragmenting v-quark, and 
 the sum on $j$ runs over all 
emitted v-gluons.
Even though the bare $q_\vg$-mass  could be as light as $10$ GeV, 
$M_{q_\vg}^{\rm eff}$ can reach values close to $M_{Q_\vg}$ 
because of radiation, as happens in QCD jets \cite{Carloni:2010tw}.
This would be especially the case for
a strongly interacting hidden/dark sector, i.e. at large $\alpha_\vg$. 
We look upon expression (\ref{eq:vgamma}) as providing an 
order of magnitude estimate of the v-quark velocity.
Of course, large variations of the $v_\hg\gamma_\hg$ factor
will occur event by event because of the wide spread of $M_{q_\vg}^{\rm 
eff}$.

In its turn, bound v-states can be formed as v-gluons create new
v-quark-antiquark pairs, as happens with gluons in a conventional QCD shower. In  
HV models with v-hadrons promptly decaying back
into SM partons, a new SM parton cascade would be originated (coexisting with invisible particles) 
eventually leading to final-state SM particles as well.
Furthermore, as the $q_\vg$ radiates more and more v-gluons, and the mean 
value 
of its effective mass $M_{Q_\vg}^{\rm eff}$ distribution shifts from 
$M_{q_\vg}$ towards $M_{Q_\vg}$, more and more energy
is subtracted from the visible quark and its associated system of emitted gluons.  In sum, a strong
coupling $\alpha_\vg$ should lead to small velocities of both SM and 
hidden particles.

Indeed, under a Lorentz boost of velocity $v_\hg$, the angular 
distribution of the final-state particles in the 
laboratory reference frame (LRF), the latter almost 
coinciding with the fragmenting $q_\vg$ 
reference frame,
is given by \cite{Byckling:1971vca}
\beq\label{eq:polar}
w(\phi-\phi_\hg)=\frac{1}{\gamma_\hg[1-v_\hg^2\cos^2{(\phi-\phi_\hg)}]}\ 
f(\phi,g)\,\, . 
\eeq 
 Here, $f(\phi,g)=(g\ \pm\ \sqrt{D})/\pm \sqrt{D}$ with
$D=1+\gamma_\hg^2(1-g^2)\tan{}^2(\phi-\phi_\hg)$, and
 $g=v_\hg/v$ 
 with  
 the final-state particle  velocity $v$ 
  in the $q_\vg$ rest frame. 
 For $g \ll 1$, one can roughly set $f(\phi,g) \approx 1$. A massive 
hidden object of spin zero as assumed for the fragmenting v-quark
in this work (leading to a nearly spherical distribution
in the $q_\vg$-quark reference frame), 
with $v_\hg$ being non-relativistic, plainly justifies
such an approximation.

\begin{figure}[ht!]
\begin{center}
\includegraphics[scale=0.61]{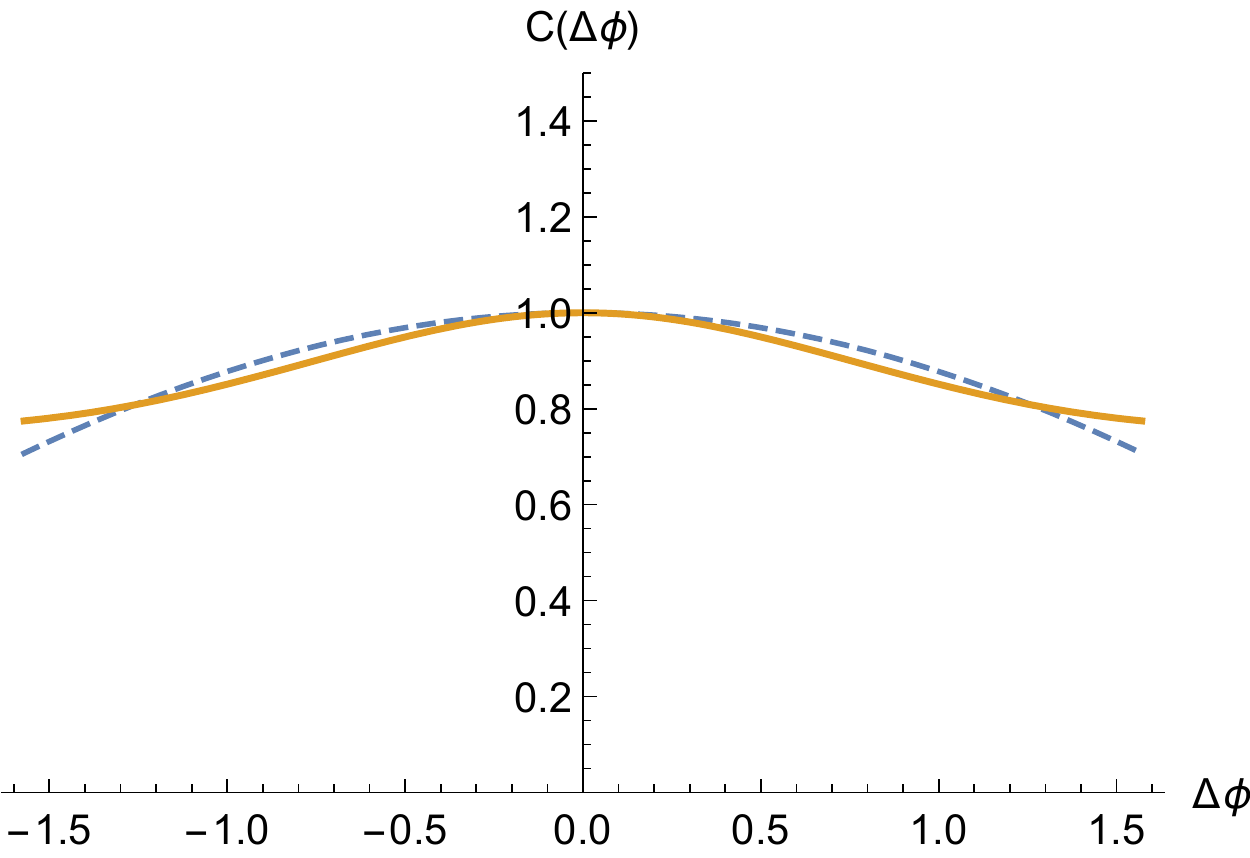}
\includegraphics[scale=0.61]{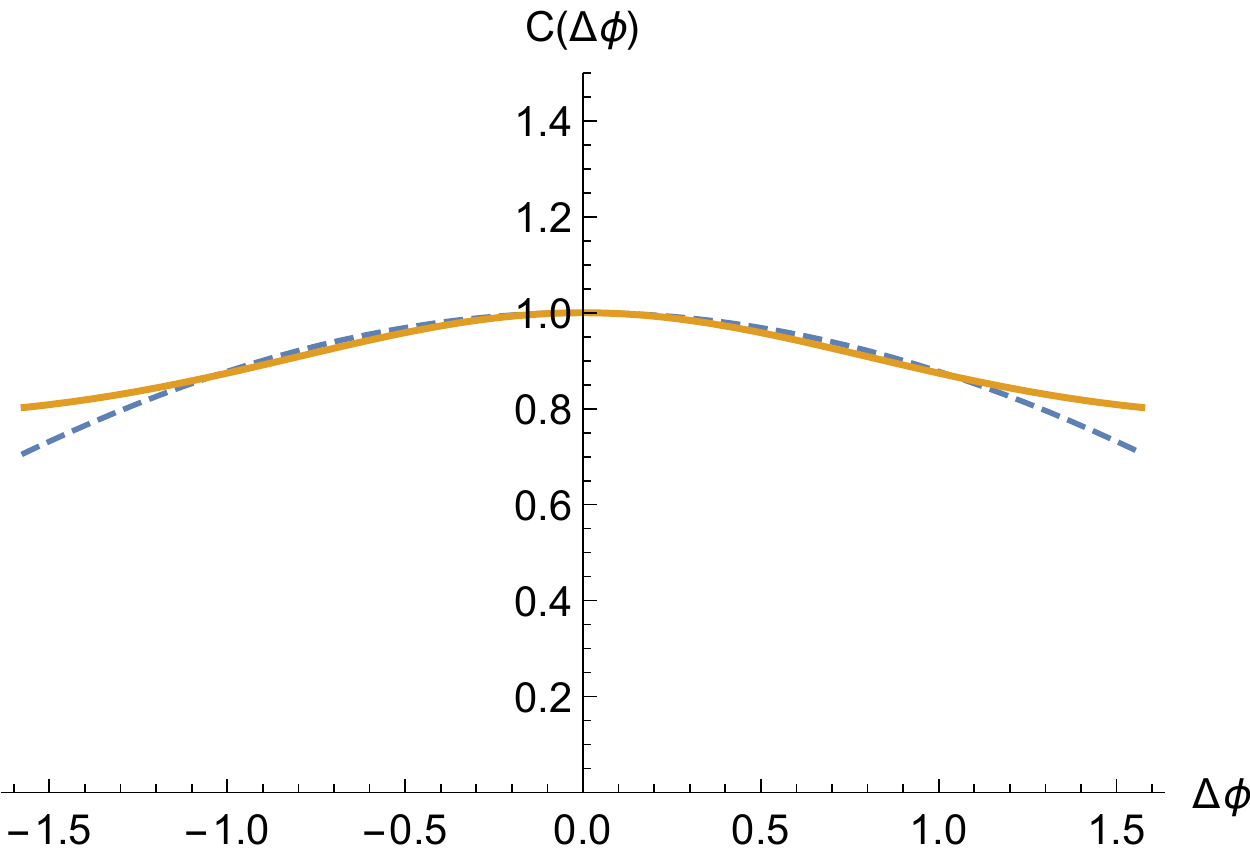}
\caption{Expected contribution to the azimuthal dependence of the 
correlation function $C(\Delta \phi)$ from a massive hidden/dark sector
(solid orange line). Left panel: $M_{Q_\vg}=300$ GeV and $M_{q_\vg}^{\rm eff}=200$ GeV. Right
 panel: $M_{Q_\vg}=1000$ GeV and $M_{q_\vg}^{\rm eff}=700$ GeV.
 A non-relativistic hidden source is considered in both cases. 
 For comparison 
 the $\cos{(\Delta \phi/2)}$ modulation (dotted blue line) is shown. The 
correlation
 function is normalized to unity at $\Delta \phi = 0$ 
 to be compared 
 with the
 $\cos{(\Delta \phi/2)}$ modulation.}
\label{fig:2}
\end{center}
\vspace*{-0.3cm}
\end{figure}

The azimuthal distribution $w(\phi-\phi_\hg)$
can be then approximated by a Gaussian for small $\phi-\phi_\hg$ angles for 
practical purposes, 
namely, 
\beq\label{eq:delta}
w(\phi-\phi_\hg)\ \approx\ 
\exp{\biggl[-\frac{(\phi-\phi_{\hg})^2}{2\delta_{\hg\phi}^2}\biggr]}\ \ \ ,\ \ 
\delta_{\hg\phi}\ \simeq\ \frac{1}{\sqrt{2}\ v_\hg\gamma_\hg} \,\, ,
\eeq
where $\delta_{\rm h\phi}$ was interpreted as an azimuthal cluster ``width''
in \cite{Sanchis-Lozano:2016qda} 
\footnote{As we are focusing on azimuthal angles, the particle trajectories are projected 
onto the transverse plane, hence the velocities $v_\hg$, $v$  and the Lorentz factor 
$\gamma_\hg$ actually correspond to transverse velocities.}.
Large hidden source velocities  
lead to small $\delta_{\hg\phi}$ and thereby
a more pronounced peak at $\phi \simeq \phi_\hg$, 
in accordance
with Eq.~(\ref{eq:polar}). 
Conversely, small velocities of the hidden source 
lead to flatter azimuthal distributions.

Substituting Eq.~(\ref{eq:vgamma})
into Eq.~(\ref{eq:delta}) for $\delta_{\hg\phi}$, one gets
\begin{equation}\label{eq:deltaM}
\delta_{\hg\phi}\ 
\simeq\ \frac{\sqrt{2}M_{Q_\vg}M_{q_\vg}^{\rm 
eff}}{M_{Q_\vg}^2-M_{q_\vg}^{{\rm 
eff}2}}\,\, ,
\end{equation}
where $M_{q_\vg}^{\rm eff}$ stands for the effective mass resulting from 
v-gluon radiation as mentionned above.

Next, by Taylor expanding the exponential we can
identify the above expression with a cosine function such that $1/\delta_{\hg\phi}$ 
determines the leading Fourier component of the NP contribution from
a given range of the effective v-quark mass.
As reference values, we set $M_{Q_\vg}=1000$ GeV
and $M_{q_\vg}^{\rm eff}=700$ GeV \cite{Carloni:2010tw}, yielding 
the closest fractional number
\begin{equation}\label{eq:integer}
\frac{1}{{\rm Integer}[\delta_{\hg\phi}]}\ =\ \frac{1}{2}\,\, .
\end{equation}
This estimate can be extended to the mass interval of the v-quark invariant mass $M_{q_\vg}^{\rm eff}\in [630,760]$ GeV, leading to the NP contribution
\beq\label{eq:cos}
 w(\phi-\phi_\hg)\ \approx\ \cos{[(\phi-\phi_\hg)/2]}\,\, 
 \eeq
from this $M_{q_\vg}^{\rm eff}$ mass ``slice''. 

By integration of the
product of the two single particle azimuthal distributions, one gets
\beq\label{eq:newterm}
C(\Delta \phi)\ \approx\ \frac{1}{2\pi}\int_0^{2\pi} \cos{[(\phi_1-\phi_\hg)/2]}\cos{[(\phi_2-\phi_\hg)/2]}\ 
d\phi_\hg\ =\ 2\ \cos{[(\phi_1-\phi_2)/2]}\,\, .
\eeq

In Fig.~\ref{fig:2} we show the expected angular
dependence of the corresponding Fourier term in the
correlation function $C(\Delta \phi)$ for two reference benchmarks:
{\em (a)} $M_{Q_\vg}=300$ GeV and $M_{q_\vg}^{\rm eff}=200$ GeV, and {\em 
(b)} $M_{Q_\vg}=1000$ GeV and $M_{q_\vg}^{\rm eff}=700$ GeV.   
 All hidden initial sources from the
 primary collision originating the subsequent visible/invisible shower
are assumed to be non-relativistic ($g$  is taken of the order of 0.1 in 
Eq.~(\ref{eq:polar})). 
 A comparison with the $\cos{(\Delta \phi/2)}$ modulation, shown at the 
 same plot,
 points  
 out
that the HS contribution should yield a Fourier component dominated by this 
term (not yet considered so far in any analysis to our knowledge).

Actually, more fractional harmonic terms should be considered as the 
whole mass range of the effective mass $M_{q_\vg}^{\rm eff}$ (up to $M_{Q_v}$)
is taken into account in the expected continous spectrum 
obtained from radiation 
\cite{Carloni:2010tw}. Hence, the Fourier series should 
be more generally written as: 
\beq\label{eq:call}
C(\Delta \phi)\ \sim 1 + 2\sum_{n=1}^{\infty}V_n\cos{(n\Delta \phi)}+
2\sum_{m=1}^{\infty}V_{1/m}'\cos{(\Delta \phi/m)}\ ,
\end{equation}
where the extra $\cos{(\Delta \phi/m)}$ harmonic terms  
encode the angular anisotropies associated with massive hidden states modifying 
the parton shower and thereby
 correlations among final-state particles. The $V'_{1/2}$ term should be
 the leading component in these fractional harmonic terms. 
 Notice 
that the
$\cos{(\Delta \phi)}$ term has now two contributions:  
a negative one ($V_1 < 0$) from
the conventional series of Eq.~(\ref{eq:cold}), and another positive one
($V_1' > 0$) expected from a HS. 

Of course, the Fourier analysis using Eq.(\ref{eq:call}) 
will contain contributions from both the conventional partonic 
cascade and from the hidden sector. 
In order to enhance such a 
hypothetical NP contribution, extra selection cuts beyond high multiplicity 
and usual $p_T$ ranges of charged hadrons
should be applied on events, in particular high $p_T$ leptons and/or missing transverse energy/momentum. 

Indeed, multi-lepton signatures have already been used in the search of 
NP at the LHC (see e.g. \cite{Aad:2014hja,Chatrchyan:2014aea})  as they are 
predicted by many models beyond the SM. For example, a cascade of partons/particles initiated by the decay of a 
heavy hidden particle can proceed though intermediate states yielding electrons, muons or tau leptons 
in the final state. Realistic requirements would imply an electron or muon with 
$p_T > 20$ GeV and $|\eta| < 2.5$, a second electron or muon with slightly looser requirements 
and a third electron, muon or hadronically decaying tau. Moreover, lepton combinations 
of the same electric charge can be used to enhance the NP signature.
Additional cuts can be large missing $E_T$, since the hidden/conventional cascade 
can result in invisible particles at the end of the decay chain. 
Lastly, as the decay of hidden particles to bottom quarks can be largely enhanced in many hidden models, 
$b$-tagging would be another technique to be applied to enrich the sample with NP events.

Notice that such proposed cuts 
 (aside a common high-multiplicity cut)
hardly could be attributed to the formation of QGP or glass condensates, 
  but 
 associated with the presence of NP. 

Thereby the sample would be enriched with NP events enhancing the ridge 
effect if due to this non-standard mechanism.
 Then the 
 non-vanishing values 
 of $V_{1/2}'$, $V_{1/3}'$ and so on,
 resulting to   
 a better fit
  than the conventional Fourier analysis,
  provide 
   a hint 
 of 
 NP,
 complementary to other kinds of searches.

Let us finally remark that, from the observation of the near-side ridge effect in hadronic collisions, an integrated luminosity of order of tens of pb$^{-1}$ 
would be needed to observe any possible NP emergent effect, provided that
the HS production cross section turns out to be large enough, crucially depending on the 
$Q_\vg$ mass \cite{Carloni:2010tw}. 
 In order to avoid pile-up effects,
 a dedicated 
 low-luminosity run would be 
 desirable 
 at the LHC.

Summarizing, hidden/dark sectors production 
on top of the parton shower in $pp$ collisions can
sizeably alter final-state particle correlations which can become a signature
of NP. Specifically, more fractional $\cos{(\Delta \phi/m)}$ harmonic terms should be included in the 
Fourier series when carrying out the analysis of
the azimuthal correlation function $C(\Delta \phi)$, once appropriate selection cuts are applied to events. 

\subsection*{Acknowledgments}
This work has been partially supported by the Spanish 
Ministerio de Ciencia, Innovaci\'on y Universidades, under grant FPA2017-84543-P, by the Severo Ochoa Excellence programme under grant SEV-2014-0398 and by Generalitat Valenciana under grant
GVPROMETEOII 2014-049. M.-A.S.L. thanks the CERN Theoretical Physics Department, where this work was started, for warm hospitality.

\end{document}